\documentclass[12pt]{article}

\usepackage{epsfig}

\def\hybrid{\topmargin -20pt  \oddsidemargin 0pt
      \headheight 0pt   \headsep 0pt
      \textwidth 6.25in 
      \textheight 9.5in 
      \marginparwidth .875in
      \parskip 5pt plus 1pt   \jot = 1.5ex}

\hybrid

\usepackage{amsfonts}
\usepackage{amssymb}

\begin{document}

\def\x{\times}
\def\pa{\partial}
\def\ra{\rightarrow}
\def\lra{\leftrightarrow}
\def\beq{\begin{equation}}
\def\eeq{\end{equation}}
\def\beqa{\begin{eqnarray}}
\def\eeqa{\end{eqnarray}}

\sloppy
\newcommand{\Tr}{{\rm Tr}}
\newcommand{\tr}{{\rm tr}}
\newcommand{\be}{\begin{equation}}
\newcommand{\ee}{\end{equation}}
\newcommand{\bea}{\begin{eqnarray}}
\newcommand{\eea}{\end{eqnarray}}
\newcommand{\eq}{\end{equation}}
\newcommand{\non}{\\ \nonumber}
\renewcommand{\arraystretch}{1.5}

\renewcommand{\thesection}{\arabic{section}}
\renewcommand{\theequation}{\thesection.\arabic{equation}}

\parindent0em

\begin{titlepage}
\begin{center}
\hfill HU-EP-01/33\\
\hfill SPIN-01/24\\
\hfill NSF-ITP-01-82\\
\hfill {\tt hep-th/0111165}\\

\vspace{1cm}

{\LARGE {\bf Fluxes and Branes in Type II Vacua  
\\
and M-theory Geometry with  
\\[.3cm]
$G_2$ and $Spin(7)$ Holonomy}}

\vspace{1cm}

{\bf Gottfried Curio},\footnote{\mbox{email}: \tt 
curio@physik.hu-berlin.de}$^{,\dag}$ 
{\bf Boris K\"ors},\footnote{\mbox{email}: \tt kors@phys.uu.nl}$^{,*}$
{\bf and Dieter L\"ust}\footnote{\mbox{email: \tt 
luest@physik.hu-berlin.de}}$^{,\dag}$
\vskip 1cm

$^\dag${\em Humboldt-Universit\"at zu Berlin,
Institut f\"ur Physik, \\ 
D-10115 Berlin, Germany} \\
\vskip .1in

$^*${\em Spinoza Institute, Utrecht University, \\
Utrecht, The Netherlands} 

\end{center}


\begin{center} {\bf ABSTRACT } \end{center}
\begin{quotation}\noindent
We discuss fluxes of RR and NSNS background fields  in type II string  
compactifications on non-compact Calabi-Yau threefolds together with 
their dual brane description which involves bound states of branes.
Simultaneously turning  on
RR and NSNS 2-form fluxes in an 1/2 supersymmetric way
can be geometrically described in M-theory 
by a $SL(2,{\mathbb Z})$ family of metrics of $G_2$ holonomy.
On the other hand, if the flux configuration
only preserves 1/4 of supersymmetries, we postulate the existence of a new
eight-dimensional 
manifold with $spin(7)$ holonomy, which does not seem to fit into
the classes of known examples.
The latter situation is dual to a 1/4 supersymmetric web 
of branes on the deformed conifold.
In addition to the 2-form fluxes, we also present some considerations on 
type IIA NSNS 4-form and 6-form fluxes.

\end{quotation}

\end{titlepage}
\vfill
\eject

\newpage

\section{Introduction}

String compactifications with background fluxes constitute an interesting
class of string vacua.
In type II or
M-theory compactifications on Calabi-Yau spaces background fluxes
are provided by vacuum expectation values of internal NSNS and RR 
H-fields [1--17]. 
These stringy fluxes cause
several very interesting effects in the effective field theory, as
they provide an effective superpotential, which may
completely or partially break the ${\cal N}=2$ supersymmetry.
E.g.
going to the rigid limit, where gravity decouples, a superpotential
with specially chosen fluxes, can realize partial supersymmetry breaking
to ${\cal N}=1$ \cite{VT}. In the corresponding four-dimensional ${\cal N}=1$
supersymmetric gauge theory this superpotential captures many interesting
effects, notably gaugino condensation, and provides a correct description
of the non-perturbative vacuum structure of these ${\cal N}=1$
gauge theories \cite{Vafa:2000wi,Cachazo:2001jy,Edelstein:2001mw}.
However for this to work, two type of stringy fluxes
are required, namely fluxes of RR as well as of NSNS $n$-form H-fields.
There is no problem in type IIB compactifications on a Calabi-Yau
space $W_6$, as the necessary fluxes are provided by internal vev's
of $H^{(3)}_R$ and $H^{(3)}_{NS}$. However in the type IIA mirror picture  
there are, at the first sight, only the RR fluxes
$H^{(n)}_R$ ($n=0,2,4,6)$ present, and the question arises what
the IIA mirror image of $H^{(3)}_{NS}$ actually is.
One of the aims of this paper is to get a better
understanding of the NS-fluxes in type IIA and their geometric interpretation 
in M-theory compactifications. 
To this end it will also be crucial to reinterpret the fluxes in terms of 
dual brane configurations. 

To be specific, consider the flux induced four-dimensional
 superpotential in type IIB
on a Calabi-Yau space $W_6$ \cite{VT,Mayr}:
\begin{equation}
W_{IIB}=\int (\tau H_{NS}^{(3)}+H_R^{(3)})\wedge\Omega
\, ,\label{supoiib}
\end{equation}
where $\Omega$ is the 
holomorphic 3-form on the Calabi-Yau space
and $\tau=a+{i/ g_s}$ is the
complex string coupling constant.
The fluxes are defined as the values of $H^{(3)}_R$ and
$\tau H^{(3)}_{NS}$, integrated over a base of 3-cycles ${\cal C}^{(3)}$.
Note that due to the contribution of 
$\tau H^{(3)}_{NS}$ these fluxes are complex,
and they contain $4h_{2,1}^{W_6}+4$ flux parameters.
The superpotential (\ref{supoiib}) transforms covariantly
under the type IIB $SL(2,\mathbb{Z})$ S-duality symmetry
with respect to $\tau$, which exchanges
$H^{(3)}_R\leftrightarrow H^{(3)}_{NS}$. 

The question is how these situations manifest itself
in the type IIA compactification on the mirror Calabi-Yau space $X_6$
with Hodge numbers $h_{1,1}^{X_6}=h_{2,1}^{W_6}$? In type IIA
the RR flux induced superpotential in four dimensions is
\cite{VT,CKKL}
\beqa
W_{IIA} = \sum_{n=0}^3\int_{X_6} H_R^{(2n)}\wedge J^{3-n}
\, .\label{supoiia}
\eeqa 
Hence the IIA superpotential is determined by the
quantized values of the 
RR H-fluxes on the
6,4,2 and 0-cycles of $X_6$, namely
\beqa
M_0=\int H_R^{(6)}, \quad M_A=\int H_R^{(4)},
\quad N_A=\int H_R^{(2)},\quad
N_0=\int H_R^{(0)}
\eeqa 
($A=1,\dots ,h_{1,1}^{X_6}$). 
Altogether the type IIA fluxes $M_I$ and $N_I$ provide
$2h_{1,1}^{X_6}+2$ real integer parameters,
and hence there is an apparent mismatch of parameters compared to the
type IIB superpotential (\ref{supoiib}). At the same time in the type IIA 
superpotential there is no dependence of the string coupling
constant $g_s$, as well as no trace of the former IIB S-duality symmetry.
 The reason for this discrepancy is of course
that in type IIA there are no even-dimensional NSNS fields strength
$H^{(2n)}_{NS}$ in the perturbative string spectrum which could
contribute to the superpotential in an obvious way.

Besides mirror symmetry, valid already in four space-time
dimensions, 
type IIA and type IIB on the same background space $X$
become T-dual to each other \cite{Dine:1989vu,Dai:1989ua,Abou-Zeid:1999fv}
if one compactifies both theories
on a circle $\mathbb{S}^1_3$ to one dimension lower
(we call this circle
$\mathbb{S}^1_3$ since in the cases of interest
$X_6$ is a six-dimensional CY-space, and therefore this circle denotes
the compact third spatial direction), i.e. starting say from
type IIA on $X_6\times \mathbb{S}^1_3$, with 
$\mathbb{S}^1_3$-radius $R_3$,  is equivalently described by
type IIB on $X_6\times \widetilde{\mathbb{S}}^1_3$, where the radius of the dual circle
$\widetilde{\mathbb{S}}^1_3$ is given by $1/R_3$. Extending this scenario
to M-theory on $X_6\times \mathbb{S}^1_{11}\times \mathbb{S}^1_3$, the type IIB S-duality
can geometrically be described by the exchange $R_3\leftrightarrow R_{11}$
\cite{Schwarz:1995dk,Schwarz:1996du,Schwarz:1996jq,Aspinwall:1996fw}, 
since the string coupling constant can
be expressed as $g_s=R_{11}/R_3$. In addition, the 
complexified coupling constant $\tau$ 
arises as the complex structure modulus of the extra torus. Due to 
this geometric M-theory
realization of S-duality in one dimension lower, we should expect
to find the trace of the needed NSNS H-fluxes in three dimensions, where
we compactify type IIA on a seven-dimensional space, respectively M-theory on
an eight-dimensional space called $X_8$. But 
in the presence of fluxes (or branes), $X_8$ will not be any longer
a direct product space $X_6\times \mathbb{S}^1_{11}\times \mathbb{S}^1_3$. 
In the following
we will examine the situation in which $X_8$ has $G_2$ or respectively  
$spin(7)$
holonomy.\footnote{For some papers on $G_2$ and $spin(7)$ manifolds see 
[28--57].} 

In the main part of the paper we will discuss the possibility of having
non-vanishing type IIA RR 2-form flux $H^{(2)}_R$ together with 
non-vanishing NSNS 2-form flux $H^{(2)}_{NS}$ through the same $\mathbb{S}^2$
inside the Calabi-Yau space $X_6$.
In M-theory these fluxes will be purely geometrical. Consider first
$N$ units of 
the flux of the RR two-form field strength
$H^{(2)}_R$, integrated over a two-sphere in $X_6$:
$N=\int_{{\mathbb P}^1}H^{(2)}_R$, but with zero NSNS 2-form flux.
Since the corresponding RR one-form gauge field
$A^{(1)}_R$, $H^{(2)}_R=dA^{(1)}_R$, originates from the eleven-dimensional
metric of M-theory, the 2-form flux is given by a certain 
topologically non-trivial metric configuration
in M-theory with topological charge $N$.
Namely the M-theory lift of the RR 2-form flux can be described by
extending the IIA 2-sphere ${\mathbb P}^1$ 
to a 3-sphere $\mathbb{S}^3$, which is modded out
by the discrete group ${\mathbb Z}_N$.
As a result, the seven-dimensional space $X_7$ is actually
not any more given by the direct product $X_6\times \mathbb{S}^1_{11}$,
where $X_6$ is a Calabi-Yau manifold, but is a circle fibration
over a six-dimensional base; in addition $X_7$
has extended holonomy,
namely, since the theory has ${\cal N}=1$ supersymmetry
in four dimensions (i.e. four conserved
supercharges), $X_7$ must have $G_2$ holonomy.

If one restricts to the local geometry, the resolved conifold, 
there is a dual description  
of the ${\cal N}=1$ large $N$ gauge theory in terms
of $N$ D6-branes which are wrapped around the 3-cycle in the deformed conifold
geometry $T^*\mathbb{S}^3$. They are BPS solutions of a suitably gauged supergravity 
with a twisted, non-commutative world volume theory \cite{Gauntlett:2001ur}.  
So the transition from 2-form flux to wrapped D6-branes
corresponds to the conifold transition in the Calabi-Yau geometry.
In M-theory, this
will be manufactured by a flop  transition in the $G_2$
manifold, which exchanges
two different 3-spheres in $X_7$.

More generally, we next consider the general situation where 
$N$ units of RR 2-form flux and $N'$ units of NSNS 2-form flux 
through the same 2-sphere are
turned on at the
same time. This means that we go from 11 to 3 dimensions on an
8-dimensional space $X_8$, which is however not any more
of the direct product form $X_6\times \mathbb{S}^1_{11}\times \mathbb{S}^1_3$,
but $X_8$ will be a torus fibration over a six-dimensional base.
Normally these fluxes also preserve 1/2 supersymmetries, and any  
R-NS flux combination can be brought back by a $SL(2,{\mathbb Z})$
transformation to a situation, where we have only RR or only NS 
2-form flux. However, as we will argue, a particular 
conspiracy of both types of 2-form fluxes can reduce the
number of conserved supersymmetries again by a factor of two, i.e.
we will deal with an eight-dimensional geometry
with two preserved supercharges which is the minimal amount of
supersymmetry in three dimensions, normally called ${\cal N}=1$
supersymmetry. In this case $X_8$ should be a manifold of
$spin(7)$ holonomy. In fact, the existence of these configurations is 
supported due to the dual brane picture in terms of 
1/4 BPS brane webs. 

More concretely, this configuration will be 
given in terms $N$ D6-branes bound together with $N'$ Kaluza-Klein
monopoles, wrapped around the 3-sphere of the deformed conifold. 
In type IIB these bound states are described by 
$(p,q)$ webs of 5-branes 
which preserve 8 supercharges before compactification. 
The three-dimensional field theory with ${\cal N}=1$ supersymmetry
will live on the three-dimensional non-compact world volume of the D6-KK system.

In the final part of the paper we want to add some remarks on how to get
NSNS 4-form flux together with RR 2-form flux, which is
necessary to describe gaugino condensation in the corresponding
large $N$ gauge theory.
As noted already in \cite{Vafa:2000wi} 
and as it can be seen from reducing down from the $G_2$ 
manifold, 
turning on RR 2-form flux on
$X_6$ implies that $X_6$ is actually not anymore a Calabi-Yau
space, but the holomorphic three-form $\Omega$ of
$X_6$ is not any more closed.
The non-closure of $\Omega$ has then the effect of a non-vanishing 
imaginary part of the NS
4-form flux \cite{Vafa:2000wi}. Again, this becomes more transparent
going to M-theory extending $X_6$ to the $G_2$
manifold. 
We will extend this discussion by arguing that a certain  4-form
can provide the
real part of the complex 4-form flux, such that the complex flux parameter
is  given by the complex, holomorphic volume of the M-theory
three-cycle.

The paper is organized as follows. In the next chapter we will discuss
some basic properties of the spectrum of M-theory on $\mathbb{T}^2$ and
the possible fluxes if one further compactifies to three dimensions.
In section three and four we will discuss the RR and NSNS 2-form fluxes 
together with the dual brane configurations and
describe their relation to eight-dimensional background spaces,
which are conjectured to possess a metric
of $spin(7)$ holonomy.
In section five we will turn to the construction of the 4-form fluxes.
We will also discuss the RR and NSNS 6-form fluxes.
The summary gives an overview over the considered cases.

\section{M-theory compactification to three dimensions}
\setcounter{equation}{0}

In this section we will discuss some generic properties of M-theory
compactified to three dimensions to set up some notation.
Let us first compactify M-theory on  a two-torus $\mathbb{T}^2$,
which can be viewed as $\mathbb{T}^2=\mathbb{S}^1_{11}\times \mathbb{S}^1_{3}$, 
or equivalently type IIA on $\mathbb{S}^1_3$ to $D=9$.
The decomposition of the metric 
yields
\be
G_{MN} \to G_{\mu\nu}, A_\mu^{3},A_\mu^{11},  \phi,\phi',a \, . 
\ee
The complex structure $\tau$ of $\mathbb{T}^2$ is built by the two scalar fields
$\phi$ and $a$,
\be
\tau=a+i\phi,\quad \phi\sim {R_3\over R_{11}}\, .
\ee
In type IIB $\tau$ is just the complexified coupling constant $\tau=a+{i/ g_s}$. 
The strong-weak coupling $SL(2,\mathbb{Z})$ transformations act on $\tau$ in the usual way:
\be
\tau\rightarrow{a\tau +b\over c\tau+d},\quad ad-bc=1\, .
\ee
Note that $A_\mu^{3}$, $A_\mu^{11}$ 
form a doublet under $SL(2,\mathbb{Z})$, which is
essentially given by the exchange of $\mathbb{S}^1_{11}$ with $\mathbb{S}^1_3$.
The decomposition of the 3-form
yields
\be
C_{MNP} \to C_{\mu\nu\rho}, B_{\mu\nu}^{3},B_{\mu\nu}^{11},
 C_{\mu} \ .\label{threeform}
\ee
In terms of NSNS and RR sectors one has
\bea
{\rm NSNS}: &&\quad  G_{\mu\nu}, B_{\mu\nu}^{3}, A_\mu^{3}, C_\mu,
\phi,\phi'\, ,\nonumber\\
{\rm RR}: &&\quad  C_{\mu\nu\rho},  B_{\mu\nu}^{11}, A_\mu^{11}, a\, .
\eea
Note that   $C_{\mu\nu\rho}, C_{\mu}$
are both singlets under $SL(2,\mathbb{Z})$.
Altogether among the odd forms we
have three 1-forms $A^{(1)}_3$, $A^{(1)}_{11}$, $C^{(1)}$
and one 3-form $C^{(3)}$.

Compactification on a six-dimensional
space $X_6$ allows $4\times h_{1,1}$ fluxes
from $F^{(2)}_{3} = dA^{(1)}_{3}$, 
$F^{(2)}_{11}=dA^{(1)}_{11}$, $G^{(2)}=dC^{(1)}$ and $G^{(4)}=dC^{(3)}$.
$F^{(2)}_{3}$ and $F^{(2)}_{11}$ form a doublet and and transform under
$SL(2,{\mathbb Z})$ as
\begin{equation}
\pmatrix{F^{(2)}_{11}
\cr F^{(2)}_3}\rightarrow
\pmatrix{a&b\cr c&d}\pmatrix{F^{(2)}_{11}\cr F^{(2)}_3}\, .\label{sltwof}
\end{equation}
Therefore they are naturally
complexified as
\beqa
H^{(2)}=H^{(2)}_R+\tau H^{(2)}_{NS}
\equiv F^{(2)}_{11} + \tau F^{(2)}_{3}.
\eeqa
This leads to a superpotential in three dimensions of the form
\be 
W = \int H^{(2)} \wedge J\wedge J=(N_A+\tau N_A')F_A\, ,\quad
F_A=\int(J\wedge J)\, .\label{suphtwo}
\ee
According to the introduction we denote the corresponding
$2\times h_{1,1}$ RR and NSNS 2-form fluxes as ($A=1,\dots ,h_{1,1}$)
\begin{equation}
N_A'=\int_{{\cal C}^{(2)}_A}H^{(2)}_{NS}=\int_{{\cal C}^{(2)}_A}F_3^{(2)},
\quad N_A=\int_{{\cal C}^{(2)}_A}H^{(2)}_R=
\int_{{\cal C}^{(2)}_A}F_{11}^{(2)}\, ,
\label{twofluxes}
\end{equation}
where ${\cal C}^{(2)}_A$ are the homology 2-cycles on $X_6$.
The 2-form fluxes will be further discussed in section 3.

For 
$C_{\mu\nu\rho}, C_{\mu}$,
being $SL(2,{\mathbb Z})$ singlets, we define a 4-form field strength 
$H^{(4)}$,
which does not include the coupling constant $\tau$, as
\beqa 
H^{(4)}=H^{(4)}_R+ H^{(4)}_{NS}\equiv
G^{(4)} +\widetilde G^{(2)},
\eeqa
where $\widetilde G^{(2)}$ is the dual of $G^{(2)}$ in $X_6$.
Hence we could also like to consider a superpotential of the form
\be 
W = \int H^{(4)} \wedge J =(M_A+M_A')X^A\, , \quad X^A=\int J\, ,
\label{supofourflux}
\ee
and call  the 
$2\times h_{1,1}$ RR and NSNS 4-form fluxes as 
\begin{equation}
M_A'=\int_{{\cal C}^{(4)}_A}H^{(4)}_{NS}=\int_{{\cal C}^{(4)}_A}\widetilde G_2,
\quad M_A=\int_{{\cal C}^{(4)}_A}H^{(4)}_R=\int_{{\cal C}^{(4)}_A} G_4\, .
\end{equation}
The ${\cal C}^{(4)}_A$ are the 4-cycles dual to the ${\cal C}^{(2)}_A$.
Further discussion of the 4-form fluxes 
and a possible geometrical origin of the $SL(2,{\mathbb Z})$
invariance can be found in section 4.

\section{NSNS and RR 2-form fluxes}
\setcounter{equation}{0}

In this chapter we present the geometrical M-theory 
interpretation of fluxes in 
type IIA 
compactifications on Calabi-Yau spaces and their dual brane settings. 
The more precise
nature of the resultant M-theory background will be further 
elucidated in chapter 4. 

\subsection{Geometric description of 2-form fluxes}

In this section we describe geometric M-theory lift of the 2-form fluxes
in a model indepedent way. Later we will embed this construction in some
particular non-compact string theory respectively M-theory background spaces.
Consider the 2-form flux through some compact 2-sphere $\mathbb{S}^2$.
This $\mathbb{S}^2$ is part of the six-dimensional space in the directions (456789),
say the $\mathbb{S}^2$ is the directions $x_4$ and $x_5$.
In addition we have the torus $\mathbb{T}^2$ in the directions 3 and 11.
Having non-trivial fluxes means that the corresponding magnetic field
strengths on $\mathbb{S}^2$ are topologically non-trivial, namely 
for one unit of flux, $A_\mu^{3}$ or $A_\mu^{11}$
essentially describes the field of a Dirac monopole;
for arbitrary fluxes $N$ or $N'$ one deals with magnetic monopoles
characterized by integer Chern numbers 
$N$ or $N'$. Since in M-theory the 1-form gauge field entirely originates
from the metric, it means that a non-vanishing flux corresponds a non-trivial
metric configuration, namely to a fibration of the M-theory torus $\mathbb{T}^2$ over the base
$\mathbb{S}^2$. We will call this four-dimensional space $\Sigma_4$.
It is finally part of the total eight-dimensional internal space, be it compact or non-compact, that locally factorizes 
\be 
X_8 = X_4 \times \Sigma_4 = X_4 \times \mathbb{P}^1 \times \mathbb{S}^1_3 \times \mathbb{S}^1_{11} . 
\ee 
Its global structure will now be characterized in various cases. 
First we concentrate on the $\mathbb{P}^1\times \mathbb{S}^1_3\times
\mathbb{S}^1_{11}$ part of $X_8$.

\vskip0.2cm
\noindent{\it Case (i): $N'=0$ or $N=0$ $\leftrightarrow$ 1/2 supersymmetry}

\vskip0.2cm

First discuss the simpler situation where either $N$ units of
RR 2-form flux or $N'$ units of NSNS 2-form flux are turned on, but not
both at the same time. The respective RR fluxes break half of the supersymmetries,
as can be seen in the corresponding dual brane picture.
In this case $\mathbb{S}^1_{11}$ is non-trivially fibred over $\mathbb{S}^2$.
For $N=1$  this is just the Hopf fibration of $\mathbb{S}^3:\ \mathbb{S}_{11}^1\rightarrow \mathbb{S}^2$.\footnote{Hopf 
fibrations, branes and fluxes were
also discussed recently in \cite{Minasian:2001sq}.}  
Due to the exchange of the two circles by the type IIB S-duality, one just has 
to change the fibre of the Hopf bundle when switching from RR to NSNS flux, $\mathbb{S}^3:\ 
\mathbb{S}_{3}^1\rightarrow \mathbb{S}^2$, for a single unit of NSNS 2-form flux. 
In the more general case of $N$ units of RR 2-form flux 
(or $N'$ units of NSNS 2-form flux)
the corresponding $\mathbb{S}^1$ fibration is the quotient of the Hopf fibration by
${\mathbb Z}_N$, and we now call this 3-dimensional
space $\Sigma_3$ ($\Sigma_3'$).
With some slight abuse of standard notation let us denote this fibration
by 
\begin{equation}
\mathbb{S}^1(N)\longrightarrow\Sigma_3\longrightarrow{\mathbb P}^1\, .
\end{equation}
The four-dimensional space $\Sigma_4$ is then given by 
\be
\Sigma_4=\Sigma_3\times \mathbb{S}^1_3 \quad {\rm or}\quad \Sigma_3'\times \mathbb{S}^1_{11}
\ee
and the total space is 
\be \label{prodspace} 
X_7 \times \mathbb{S}_3^1\quad {\rm or}\quad X'_7 \times \mathbb{S}_{11}^1 .
\ee
Here, $X_7$ and $X'_7$ have as holonomy the group $G_2$ \cite{Atiyah:2000zz}. 

\vskip0.2cm
\noindent{\it Case (ii): $N,N'\neq0$ $\leftrightarrow$ 1/2 supersymmetry}

\vskip0.2cm

Now assume that we have non-zero RR and NSNS 2-form flux through
the same $\mathbb{S}^2$, i.e.
$N$ and $N'$ are both non-zero. 
Normally, these fluxes
also preserve 1/2 of supersymmetries. 
This can be seen from the fact that any flux configuration with quantum numbers
$(N\ N')^T$ 
($N$, $N'$ relatively prime) 
can be transformed by a $SL(2,\mathbb{Z})$ transformation
to the flux vector $(1\ 0)^T$:
\begin{equation}
\pmatrix{N\cr N'}=\pmatrix{N&b\cr N'&d}\pmatrix{1\cr 0}\, .\label{sltwo}
\end{equation}
Analogously, a general flux vector $(N\ N')^T$ with $n={\rm gcd}(N,N')$ 
is related to $(n\ 0)^T$ via an $SL(2,\mathbb{Z})$ transformation.

In the first place, the whole torus $\mathbb{T}^2=\mathbb{S}^1_3\times \mathbb{S}^1_{11}$ appears 
non-trivially fibred over $\mathbb{S}^2$, leading to the four-dimensional space 
$\Sigma_4$. But, according to the above transformation of the fluxes, there is a linear 
combination of the two circles that is trivially fibered, while its dual circle 
is fibered non-trivially and with regard to $n={\rm gcd}(N,N')$ units of flux. 
In other words one starts with ${\mathbb P}^1$ as base space and 
a two-dimensional $\mathbb{S}^1_3\times \mathbb{S}^1_{11}$ fibre.
In the simplest case $N=N'=1$, this $\Sigma_4$ is a `Hopf fibre product', namely a
$\mathbb{T}^2=\mathbb{S}^{1}_3\times \mathbb{S}^{1}_{11}$ fibration, over ${\mathbb P}^1$, 
which we denote by
\begin{equation} 
\mathbb{S}_{11}^1\longrightarrow {\mathbb P}^1\longleftarrow \mathbb{S}_3^1\, .
\end{equation}
However, the torus fibration is such that by acting on the complex structure
$\tau$ of the torus by the following $SL(2,\mathbb{Z})$ transformation,
\begin{equation}
\tau\rightarrow {\tau\over\tau+1}\, ,
\end{equation}
in the new homology basis one $\mathbb{S}^1$ is trivially fibred over $\mathbb{S}^2$, whereas the other 
$\mathbb{S}^1$ is the fibre of a Hopf bundle whose total space is an $\mathbb{S}^3$.
This is the precise geometric analogue of the fact
that the flux vector $(1\ 1)^T$ is mapped to the flux vector $(1\ 0)^T$.
Therefore the space $\Sigma_4$ is topologically again given by
$\mathbb{S}^3\times \mathbb{S}^1$. 

More generally, this space is divided by a discrete product group
${\mathbb Z}_N\times {\mathbb Z}_{N'}$,
where ${\mathbb Z}_N\subset \mathbb{S}^1_{11}$ and ${\mathbb Z}_{N'}\subset
\mathbb{S}^1_3$. 
This means that $\Sigma_4$ is a $\mathbb{T}^2$ fibration, namely a fibre product,
characterized by two winding numbers $N$ and $N'$.
We denote the fibre product $\Sigma_4$ similar as before as
\begin{equation} 
\Sigma_4:
\qquad {S}^1(N)\longrightarrow {\mathbb P}^1\longleftarrow {S}^1(N')\, .
\end{equation}
But as before, acting on $\tau$ with the same $SL(2,\mathbb{Z})$
transformation as in 
eq.(\ref{sltwo}) one circle is trivially fibred, and $\Sigma_4$ is given as
\be
\Sigma_4=\Sigma_3\times \mathbb{S}^1\, .
\ee
Thus we see, that both types of 1/2 BPS configurations are related to a geometry where 
the total eight-dimensional space is still of a product structure of the kind (\ref{prodspace}) 
and thus preserves more than the minimal supersymmetry in $d=3$. \\ 

Starting from the metric of $\mathbb{S}^3$,
the metric of $\Sigma_4$ can be constructed in a straightforward way.
Specifically parametrize $\mathbb{S}^3$ in terms of the three Euler angles
$\theta$, $\phi$ and $\psi$. The coordinates $\theta$ and $\phi$
are the coordinates of the $\mathbb{S}^2$ and $\psi$ corresponds to the
Hopf fibre $\mathbb{S}^1$. Then the standard metric of $\mathbb{S}^3\times \mathbb{S}^1$, 
which corresponds to the flux vector $(1\ 0)^T$,
has the form
\begin{equation}
ds^2=(\sigma_1)^2+(\sigma_2)^2+(\sigma_3)^2+(d\psi')^2\, ,
\end{equation}
where $\psi'$ is the coordinate of the additional circle, and
the $\sigma_a$ are the left-invariant 1-forms on $SU(2)$:
\begin{equation}
\label{sigmas}
\sigma_1=\cos\psi ~d\theta+\sin\psi\sin\theta ~d\phi ,\quad
\sigma_2=-\sin\psi ~d\theta+\cos\psi\sin\theta ~d\phi ,\quad
\sigma_3=d\psi+\cos\theta d\phi\, .
\end{equation}
Now consider the $SL(2,\mathbb{Z})$ transformation in eq.(\ref{sltwo})
which transforms to the flux vector $(N\ N')^T$.
Acting with this transformation on the complex structure $\tau$ of
the torus is equivalent to perform a holomorphic reparametrization
on the torus coordinates, which is just a linear 
transformation on the real coordinates 
$\psi$ and $\psi'$ of the form
\begin{equation}
\pmatrix{\psi\cr \psi'}\rightarrow
\pmatrix{N&b\cr N'&d}\pmatrix{\psi\cr \psi'}\, .\label{sltwopsi}
\end{equation}
Then $\Sigma_4$ is described by an $SL(2,{\mathbb Z})$ family
of metrics labeled by the two integers $N,N'$ is the following way:
by 
\begin{equation}
\Sigma_4:\qquad ds^2=(\sigma_1')^2+(\sigma_2')^2+(\sigma_3')^2
+(\sigma_4')^2\, .
\label{metricsigmafour}
\end{equation}
Now the $\sigma_a'$ take the form:
\beqa
\sigma_1'&=&\cos(N\psi+b\psi') ~d\theta+\sin(N\psi+b\psi')
\sin\theta ~d\phi ,\nonumber\\
\sigma_2'&=&-\sin(N\psi+b\psi')~ d\theta+\cos(N\psi+b\psi')
\sin\theta ~d\phi ,\nonumber\\
\sigma_3'&=&Nd\psi+bd\psi'+\cos\theta ~d\phi , \nonumber\\
\sigma_4'&=&N'd\psi+d(d\psi') \, .
\label{sigmaprimes}
\eeqa 
In this way, the $SL(2,\mathbb{Z})$ that acts on the charge lattice extends to 
the metric on $\mathbb{S}^3\times \mathbb{S}^1$.
We now turn to configurations where the local geometry breaks to the minimal 
two supercharges in $d=3$, i.e. 
should be given by a $spin(7)$ instead of a $G_2$ manifold. 

\vskip0.2cm
\noindent{\it Case (iii): $N,N'\neq0$ $\leftrightarrow$ 1/4 supersymmetry}

\vskip0.2cm
In the following we will discuss a torus fibration 
${\mathbb T}^2\longrightarrow\Sigma_4\longrightarrow{\mathbb P}^1$
which preserves 1/4 instead of 1/2 supersymmetry, which will actually  
be justified by referring to its dual brane realization later on. 
We require
that the fibration is such that via a $SL(2,\mathbb{Z})$ transformation
one cannot transform back to the situation where one cycle of the torus
is trivially fibred over $\mathbb{S}^2$. In this way the NSNS and RR gauge fields
are
non-trivially intertwined.  We will geometrically realize this situation
by demanding the fibration is degenerate over some points of the $\mathbb{S}^2$,
where one or both cycles of the torus shrink to zero size. There is no 
$SL(2,\mathbb{Z})$ transformation which maps this flux configuration back to any single 
2-form flux vector $(N\ 0)^T$, such that the topology of the four-dimensional
space $\Sigma_4$ does no longer factorizes through $\Sigma_3\times \mathbb{S}^1$.
In terms of its linear action on the coordinates of the $\mathbb{T}_{3,11}^2$, 
this can be rephrased in 
the statement that the $SL(2,\mathbb{Z})$ equivalence breaks down as soon as 
cycles degenerate. In a similar situation, 
the related absence of a symplectic basis in an effective ${\cal N}=2$ field theory 
in $d=4$ was recognized as a prerequesite for a partial breaking of 
supersymmetry via 
fluxes on Calabi-Yau spaces in type IIA \cite{VT}.  

As a specific example of a torus fibration with degenerate fibres let us
discuss the space ${\mathbb CP}^2$.
Later, we will come to the point that ${\mathbb CP}^2$ appears in an
eight-dimensional $spin(7)$ manifold based on the $SU(3)/U(1)$ coset. 
So let us briefly recall the toric construction of ${\mathbb CP}^2$
\cite{Leung:1998tw}.
As it is well known, ${\mathbb CP}^2$ can be represented as
\begin{equation}
\left( {\mathbb C}^3\backslash\lbrace 0\rbrace \right) /{\mathbb C}^\star=\mathbb{S}^5/U(1)\, .
\end{equation}
The fibre torus is defined by the $U(1)^2$ action, which is given by the
$U(1)^3$ action on the three complex coordinates $z_i$, modulo
the action of the diagonal $U(1)$. 
The basis of the $U(1)^2$ action can be chosen as
\begin{equation}
(z_1,z_2,z_3)\rightarrow (z_1\exp(i\psi),z_2\exp(i\psi'),z_3)\, .
\end{equation}
The action of the two respective 
$U(1)$'s has as fixed loci two ${\mathbb P}^1$'s parametrized
by $z_2/z_3$ and $z_1/z_3$ respectively. 
This means that we can view ${\mathbb CP}^2$ as a $\mathbb{T}^2$ fibration with
generic $\mathbb{T}^2$ fibres with coordinates $(\psi,\psi')$ and with base
parametrized by $r_1=|z_1/z_3|$ and $r_2=|z_2/z_3|$.
This base has the form of a triangle, where the $\mathbb{T}^2$
fibration degenerates at the three edges of this triangle.
More specifically, over one edge the $a$ homology cycle of $\mathbb{T}^2$ is
shrinking to zero size, at the second edge the $b$ cycle degenerate,
at the third edge the $a+b$ cycle is zero, and at the three vertices
the whole torus degenerates. 

Clearly, the description of ${\mathbb CP}^2$ as $\mathbb{T}^2$ fibration over
the triangle is not quite adapted to what we have in mind: The base
triangle is a manifold with boundary and as such not suitable to lead
to an integrated flux and the corresponding element of the second
cohomology; among other things the independence of the flux value of
the (to be integrated over) cycle in its homology class assumes the
cycle boundaryless (the difference of two representatives rather
should be the boundary of another cycle so that the relevant 
field strength vanishes after integration over that difference);
clearly, in case the representatives have boundaries themselves,
the boundary of the potential 'connecting' cycle (of one dimension
higher) will consist of further pieces.

So instead of ${\mathbb CP}^2$ we will rather
describe a $\mathbb{T}^2$ fibration over an $\mathbb{S}^2$. For this
note that with $z_k=r_ke^{i\phi_k}$ the $\mathbb{S}^5$ in 
${\mathbb CP}^2=\mathbb{S}^5/U(1)_D$ has the description 
$\sum_{k=1}^3r_k^2e^{2i\phi_k}=1$. Clearly this restricts the $z_k$ to
their respective disks $|z_k|\leq 1$. Here $r_k$ runs
over $[0, 1]$ (with $\phi_k\in [0,2\pi ]$
where the endpoints are identified). Now let 
run $\phi_k$ instead rather only over $[0,\pi ]$
with the endpoints identified and $r_k$ over $[-1, +1]$ (so that
positive/negative $r_k$ covers the upper/lower half-disk). That is, we
operate with the factor group $\widetilde{U(1)_k}:=U(1)_k/{\mathbb{Z}_2}$ with
${\mathbb{Z}_2}=\{ \pm 1\}$. Now we will factor the projection
$\mathbb{S}^5\rightarrow {\mathbb CP}^2=\mathbb{S}^5/U(1)_D$ through 
$\pi: \mathbb{S}^5\rightarrow \widetilde{{\mathbb CP}^2}=\mathbb{S}^5/\widetilde{U(1)_D}$
where $\widetilde{{\mathbb CP}^2}\rightarrow {\mathbb CP}^2$ is a ${\mathbb{Z}
_2}$ cover. Then
the other projection $\mathbb{S}^5\rightarrow \mathbb{S}^2$, with the $\mathbb{S}^2$ given by 
$\sum_{k=1}^3r_k^2=1$, which factors out
$\prod_{k=1}^3\widetilde{U(1)_k}$, will factor over $\pi$, i.e.
one has a $\mathbb{T}^2=\prod_{k=1}^3\widetilde{U(1)_k}/\widetilde{U(1)_D}$ fibration
of $\widetilde{{\mathbb CP}^2}$ over the mentioned $\mathbb{S}^2$.
As a check note that the fibre gives a non-trivial contribution (of
$+1$) to the Euler number of $\widetilde{{\mathbb CP}^2}$ only if it does not
contain {\em any} $\mathbb{S}^1$ any longer (`complete degeneration'); this
occurs over the six points $r_k=\pm 1$ (and the other two $r_i$'s
zero) for $k=1,2,3$ leading to 
$e(\widetilde{{\mathbb CP}^2})=6$ as it should be for this two-fold cover
of ${\mathbb CP}^2$.


This structure as a degenerate fibration of $\mathbb{C}P^2$ is 
also evident in the Fubini-Study metric. 
The corresponding K\"ahler potential is
\beqa
\label{Kaehl pot}
K=\log (|z_1|^2+|z_2|^2+|z_3|^2)
\eeqa
and from the associated (1,1) form $\omega=i\pa \bar{\pa} K$
one gets the metric in the patch $z_3=1$
(with $z:=z_1, w:=z_2$; note that $wdz-zdw=w^2d(z/w)$)
\beqa
\label{Fub Stu metr}
ds_{\rm FS}^2=\frac{1}{(1+z\bar{z}+w\bar{w})^2}
\Bigl( dz d\bar{z} + dw d\bar{w} 
+ (wdz-zdw) (\bar{w}d\bar{z}-\bar{z}d\bar{w})\Bigr) . 
\eeqa
To connect to the description as a degenerate fibration of a 
$\mathbb{T}^2$ over a $\mathbb{P}^1$ 
note first that (\ref{Kaehl pot}) is 
invariant under the $SU(3)$ action; the $\mathbb{S}^5=\{K(z_i)=0\}$ is of course the one
in $ \mathbb{S}^5 \cong SU(3)/SU(2)$ which underlies the isomorphism
$SU(3)/U(2)\cong {\mathbb CP}^2$; the patches in $\mathbb{S}^5$ are the complements of
the three sets $\{z_i=0\} \subset \mathbb{S}^5$. 
One can then switch to polar coordinates 
in the patch ${\mathbb{C}^2_{z,w}}\subset {\mathbb CP}^2$. With
\beqa
z=R \cos 4\phi \exp\left({i\frac {\psi + \psi'}{2}}\right) \;\;\; , \;\;\;
w=R \sin 4\phi \exp\left({i\frac {\psi - \psi'}{2}}\right)\, , 
\eeqa
which identifies $\psi\pm \psi'$ as the coordinates of the fibre torus 
and $R$ and $\phi$ as those of 
the base. 
In order to find a natural splitting of the metric into fibre and base   
it is more convenient to go back to the definition of 
$\widetilde{\mathbb{C}P^2}$ above. 
In the patch $z_3\not=0$ we can switch to coordinates
\be \label{pqcoord} 
\Big[ \frac{z}{\sqrt{1+z\bar z + w\bar w}}, \frac{w}{\sqrt{1+z\bar z + w\bar w}} ,
\frac{1}{\sqrt{1+z\bar z + w\bar w}} \Big] = 
\Big[ p,q,\sqrt{1-|p|^2 -|q|^2} \Big] . 
\ee
The fibre torus is then given by the phases of $p$ and $q$, 
while the base space is 
parametrized by $|p|$ and $|q|$ subject to $|p|^2+|q|^2\le 1$ within $[0,1]$. 
Extending their domain to $[-1,1]$ and including as 
well the opposite sign for the square root in (\ref{pqcoord}) introduces the 
transition from $\mathbb{C}P^2$ to its double covering space, extending the base from a disc to 
a sphere $\mathbb{S}^2 \subset \mathbb{R}^3$.  
Evidently, the fibre degenerates at the poles of the $\mathbb{S}^2$ where any one of the coordinates 
$p$ and $q$ vanishes, or both. Note that this construction of a (degenerate) 
torus fibration is no longer elliptic but a purely real object. 
We can now relate the standard coordinates $\theta\in [-\pi/2,\pi/2]$ and 
$\phi\in [0,{\pi\over 2} ]$ on $\mathbb{S}^2$ to $p,q$ and $z,w$. Defining
\be
\cos \theta = \sqrt{|p|^2+|q|^2},  
\quad \cos 4\phi = \frac{|p|}{\sqrt{|p|^2+|q|^2}} ,  
\ee
the Fubini metric (\ref{Fub Stu metr}) on $\widetilde{\mathbb{C}P^2}$ translates into 
\be \label{splitting}
ds_{\rm FS}^2 = \frac{1}{4} \left( \sin^2 \theta\ (\sigma_1^2 + \sigma_2^2 ) + 
\sin^2 \theta \cos^2 \theta\ \sigma_3^2 \right) + d\theta^2 , 
\ee
where the left-invariant $SU(2)$ differentials $\sigma_i$ read
\be 
\sigma_1^2 + \sigma_2^2 =  d\phi^2 + \sin^2 (8\phi)\ (d\psi')^2 , \quad 
\sigma_3^2 = d\psi^2 + 2\cos (8\phi)\ d\psi d\psi' + \cos^2 (8\phi)\ (d\psi')^2 . 
\ee
In order to exhibit the torus fibration structure with base coordinates
$\theta,\phi$ and fibre coordinates $\psi,\psi'$ a different splitting
may also be convenient:
\beqa \label{fubini} 
ds_{\rm FS}^2
=\widetilde\Sigma_1^2+\widetilde\Sigma_2^2+
\widetilde\nu_1^2+\widetilde\nu_2^2 , 
\eeqa
where
$\widetilde\Sigma$ and $\widetilde\nu$ are identified as 
\beqa \label{tildediff}
\widetilde\Sigma = d\theta + i \sin\theta\ d\phi , \quad 
\widetilde\nu = 
\frac{1}{2} \sin \theta \sin(8\phi)\ d\psi' + \frac{i}{2} 
\sin\theta \cos\theta \left( 
d\psi + \cos(8\phi) d\psi' \right) .
\eeqa

One can now readily reproduce the limiting cases when one or both of the fibre circles degenerate: 
When $\phi$ goes to either 0 or $\pi/8$,  
then $\sigma_3 \longrightarrow d\psi \pm d\psi'$ while 
$\sigma_1^2 + \sigma_2^2 \longrightarrow d\phi^2$. 
Then the metric reduces to 
\be
ds_{\rm FS}^2(\phi=0,{\pi\over 8}) 
= d\theta^2 + \sin^2\theta\ d\phi^2 + 
\sin^2\theta \cos^2\theta\ d\left( \psi \pm \psi' \right)^2 .
\ee
This is the metric induced by the flat metric on $\mathbb{R}^3$, the standard metric on the 
$\mathbb{S}^2$, times a circle which is fibered over a line segment 
$\theta \in [-\pi /2,\pi/2]$ and 
vanishes at the degeneration points, where $z,w=0$ or $\infty$ simultaneously.  
This metric, of course, does no longer describe an $\mathbb{S}^3$, just as the 
given differentials do not 
satisfy an $SU(2)$ algebra. 

The structure of degeneration in the fibre-torus does not 
allow a trivialization of any 
$\mathbb{S}^1$ via the $SL(2,\mathbb{Z})$: 
Within the smooth domain the geometry can be mapped to that of 
a single flux vector. 
But along the lines, where any single $S^1$
degenerates and
at the 
intersection locus, where the two of the three cycles 
$a$, $b$ and $a+b$ degenerate simultaneously,
no such equivalence exists. 
Instead, one can map to the `standard' 
configuration, where just the circles $\mathbb{S}^1_3$, $\mathbb{S}^1_{11}$ and a  
(homological) linear combination of the two degenerate along the 
respective lines in the base.   
This will become evident also in the dual brane picture by identifying the 
degeneration locus with  
a $(p,q)$ web of branes, which cannot be trivialized at the intersections 
either. 

In fact, the description of the geometry translates into 
a more precise statement for the corresponding
fluxes in the effective type II theory. 
Along the smooth domain, where the torus is non-degenerate, the fibration
can be entirely described by RR plus NSNS 2-form fluxes.
However at the singular locus, besides the fluxes also some branes
are needed. Namely the singular loci
will be interpreted as the loci of KK monopoles
respectively  D6-branes. Thus we are lead to 
consider that the description of the resolved conifold still 
involves branes along circles inside the base $\mathbb{P}^1$, in 
addition to the 
fluxes present anyway. 
For example, the intersection locus $L$, where both cycles of the torus
degenerate, is of codimension four in $X_8$. In this case we
have additional D6-branes, located at $L$, and with world volumes
along ${\mathbb R}^{1,2}\times {\cal C}^{(4)}$, where ${\mathbb R}^{1,2}$
is the uncompactifies 3-dimensional space time, and ${\cal C}^{(4)}$
is some (non-compact) cycle inside $X_8$.
Nevertheless, the configuration is still of completely geometrical nature 
in eleven dimensions, as only D6-branes of type IIA are involved in the brane 
configuration, which we turn to now.

\subsection{The dual brane picture of the 2-form fluxes}

We shall now construct the relevant brane configurations in 
type IIA and IIB, which are dual 
to the compactifications with background fluxes discussed above. 
Again, we proceed 
case by case.  

\vskip0.2cm
\noindent{\it Case (i): $N'=0$ or $N=0$ $\leftrightarrow$ 1/2 supersymmetry}

\vskip0.2cm

The dual IIA brane description of the 
RR 2-form fluxes is given by $N$ D6-branes (being wrapped around
an $\mathbb{S}^3$ in the direction
456). Like $N$ units of RR flux, the $N$ wrapped D6-branes preserve
1/2 of the supersymmetries, when wrapping around a sLag cycle. The M-theory
lift of the D6 branes is indeed purely geometrical, namely they are given
by $N$ coincident Kaluza-Klein (${\cal KK}$) monopoles. 
These are given in terms of an $A_{N-1}$ singularity filling
the 789(11) spatial directions, where the
isometry direction of the ${\cal KK}$ monopoles is to be indentified with
the M-theory circle $\mathbb{S}^1_{11}$.
For $N=1$ the $A_1$ singularity is just given by the space
$T^*{\mathbb P}^1$ \cite{Leung:1998tw}; modding out this space
by the $U(1)$ circle action in the eleventh direction, $T^*{\mathbb P}^1/U(1)$
describes the 3-dimenensional IIA geometry.
The two degeneration points on this space, where the $U(1)$ fibre
has zero size, corresponds to the locations of the two D6-branes.
Analoguously, any NS-flux is then geometrically described by flipping the two circles and 
applying the above accordingly. 

\vskip0.2cm
\noindent{\it Case (ii): $N,N'\neq0$ $\leftrightarrow$ 1/2 supersymmetry}

\vskip0.2cm

The dual brane picture of $N$ units of RR 2-form flux
together with $N'$ units of NSNS 2-form flux reveals a better understanding for 
the amount of supersymmetry preserved by the fluxes: Recall that in type IIA 
the RR 2-form flux corresponds to $N$ D6-branes wrapped around
an $\mathbb{S}^3$, where 
the D6-brane world volumes are for example in the directions
0123456 (456 are the compact $\mathbb{S}^3$ directions).
Adding the NSNS 2-form flux means that the dual brane picture corresponds to
a boundstate of $N$ D6-branes together with $N'$ type IIA KK monopoles
which extend in the directions 012456, as depicted in table \ref{boundstatea} (the x denote the
world volume directions, and the $\cdot$ correspond to the
isometry directions of the KK (${\cal KK}$) monopoles). 

\parbox{\textwidth}
{
 \refstepcounter{table}
 \label{boundstatea}
 \begin{center}
 \begin{tabular}{|c|ccccccccc|}
\hline
                      & 1 & 2 & 3  &  4  &5&6&7&8&9      \\[-1.75ex]
  D6                  & x & x & x & x & x &x & & &    \\
  KK                  & x & x & $\cdot$ & x  & x&x& & &      \\[1ex] 
 \hline
 \end{tabular} 
 \end{center}
 \center{{
 Table~{\thetable}.} IIA D6-KK  bound state configuration}
}
\\[2ex]

The circle $\mathbb{S}^1_3$ is the isometry direction of the KK monopoles.
Going to type IIB via a T-duality along $\mathbb{S}^1_3$, one obtains
a $(N,N')$ 5-brane configuration, namely a non-threshold boundstate of
$N$ D5-branes with $N'$ NS5-branes \cite{Schwarz:1995dk,Witten:1996im}, 
with world volumes both in the
directions 012456, which means they are wrapped around the same $\mathbb{S}^3$, as given in 
table \ref{boundstatem}.

\parbox{\textwidth}
{
 \refstepcounter{table}
 \label{boundstatem}
 \begin{center}
 \begin{tabular}{|c|ccccccccc|}
\hline
                      & 1 & 2 & 3  &  4  &5&6&7&8&9      \\[-1.75ex]
  D5                  & x & x &  & x & x &x & & &    \\
  NS5                  & x & x &  & x  & x&x& & &      \\[1ex] 
 \hline
 \end{tabular} 
 \end{center}
 \center{{
 Table~{\thetable}.} IIB D5-NS5  bound state configuration}
}
\\[2ex]

This boundstate is analogous to the $(N,N')$ string, which is a
boundstate of $N$ elementary strings and $N'$ D1-branes.
Via $SL(2,\mathbb{Z})$ transformations the $(N,N')$ 5-brane 
can be related  to a 5-brane state with quantum numbers
$(n,0)$. Therefore the non-threshold $(N,N')$ 5-brane preserves 1/2
supersymmetry. Alternatively one can start with a threshold bound state
of $N$ D5-branes together with $N'$ NS5-branes which
are parallel, i.e. they share the same world volume directions.  Then 
the Killing spinor equations,
\beqa
{\rm NS}5: \quad &{~}&  \epsilon_L=\Gamma_0 \Gamma_1\dots \Gamma_5 \epsilon_L, \quad
    \epsilon_R=- \Gamma_0 \Gamma_1\dots\Gamma_5 \epsilon_R\, , \nonumber\cr
{\rm D}5:\quad&{~}&       \epsilon_L=\Gamma_0 \Gamma_1 \dots \Gamma_5 \epsilon_R \, , 
\eeqa
cannot be satisfied at the same time and no supersymmetry is preserved.
However the condensation to a 1/2 BPS non-threshold boundstate can be
thought of as the result of switching on some gauge field flux on the
NS5-brane just as the (1,1) string can be thought of as the result of
switching on electric flux on the D-string. 

The M-theory lift of the type IIA situation is described in terms
of a configuration of $N$ ${\cal KK}$ 
monopoles with world volumes along 0123456 and isometry in $\mathbb{S}^1_{11}$
together with $N'$ ${\cal KK}'$ monopoles along 012456(11) and isometry in
$\mathbb{S}^1_3$. This is summarized in table \ref{boundstate}. 

\parbox{\textwidth}
{
 \refstepcounter{table}
 \label{boundstate}
 \begin{center}
 \begin{tabular}{|c|cccccccccc|}
\hline
                      & 1 & 2 & 3  &  4  &5&6&7&8&9 &11     \\[-1.75ex]
  ${\cal KK}$                  & x & x & x & x & x &x & & & &$\cdot$    \\
 ${\cal KK}'$                  & x & x & $\cdot$ & x  & x&x& & &  & x  \\[1ex] 
 \hline
 \end{tabular} 
 \end{center}
 \center{{
 Table~{\thetable}.} M-theory ${\cal KK}$-${\cal KK}'$ bound state 
 configuration}
}
\\[2ex]

There is a very direct way to deduce the same results also directly within 
type IIA. One can view the configuration of table \ref{boundstate} as a 
configuration of D6-branes upon compactifying along any of the 789 directions, 
for instance. 
Another T-duality along 3 or 11 then produces a D$(2p)$-D$(2p-2)$ ($p=3$) bound state, 
which where 
discussed 
e.g. in 
\cite{Mihailescu:2001dn,Witten:2000mf,Blumenhagen:2000eb}.\footnote{See 
also the 
generalization in \cite{Kors:2001fh} 
for bound states with additional fluxes turned on, where relations to 
non-commutative geometry were drawn.} 
The open string spectrum contains a bifundamental tachyon that signals the 
condensation towards the non-threshold bound state. 
At a certain value for the background gauge field on the brane, the tachyon 
becomes massless and 
the deformation between the bound state and the 
superposition of the two branes with this flux is marginal. 
In any case, the bound state of a  D$(2p)$-D$(2p-2)$ brane system is 1/2 BPS, 
which confirms the result for the type IIB picture.  
The endpoint of the tachyon condensation in this simplest case is given by a 
supersymmetric 
1-cycle of appropriate RR charge, i.e. $N$ D6-branes along $\mathbb{S}^1_3$ and $N'$ 
D6-branes along 
$\mathbb{S}^1_{11}$ will condense into a single stack of $n={\rm gcd}(N,N')$ D6-branes along the cycle 
with homology class $(N/n,N'/n)$ in terms of the classes of the two circles.    
Thus, the condensation of branes in type IIA provides a geometric realization of the charge 
lattice in type IIB, which has its origin in the homology lattice of the torus. 

By a similar reasoning one can also infer that no such bound state of D$(2p)$-D$0$ branes 
can give rise to a local geometry of $spin(7)$ holonomy. This would imply a 1/16 BPS bound 
state, given by a D8-D0 superposition. Upon T-dualities, this cannot be mapped to a 
configuration of D6-branes, but instead to D5-branes, just by counting the number of 
duality transformations necessary, 4 along and 1 transverse to the D8-branes.  
Thus, this state cannot be exclusively geometric in M-theory.  

\vskip0.2cm
\noindent{\it Case (iii): $N,N'\neq0$ $\leftrightarrow$ 1/4 supersymmetry}

\vskip0.2cm

Now we are dealing with a $(N,N')$ boundstate
of two Kaluza-Klein monopoles in M-theory, which preserves 1/4 of supersymmetries.
For convenience let us discuss the 1/4 BPS state in the more familiar,
T-dual type IIB picture, where it maps to an $(N,N')$ web
of 5-branes of type IIB \cite{Aharony:1997ju,Aharony:1998bh}.
The local singularity is now given by a six-dimensional, non-compact
Calabi-Yau space, which in the most simple case can be described as 
$N({\mathbb P}^2)$ \cite{Leung:1998tw} (say in the directions 36789 and 11)
which denotes ${\mathbb CP}^2$ (or according
to our previous discussion perhaps better its 2-fold cover
$\widetilde{{\mathbb CP}^2}$) with the normal bundle on top of it.
This simplest geometry refers to the presence of a (1,0) brane, a (0,1) brane and, for the 
sake of conserving charge at the common intersection, a (-1,-1) brane. 
Each brane taken separately can be well described in terms of RR or NSNS 2-form fluxes, 
respectively by a fibration of the appropriate $\mathbb{S}^1$ over the base $\mathbb{P}^1$. 
Also linear combinations of these fluxes could be reduced to the case of just 
a single one in 
the manner described in the previous section by employing the $SL(2,\mathbb{Z})$. But 
the non-threshold brane web setting is not equivalent to any single 2-form flux vector.  

This space then replaces $\Sigma_4=\widetilde{{\mathbb CP}^2}$ in the dual flux picture.
$N({\mathbb CP}^2)$ inherits a $\mathbb{T}^2$ fibration from ${\mathbb CP}^2$;
the four-dimensional space $N({\mathbb C}P^2)/\mathbb{T}^2_{3,11}$ is trivial
except on the `triangle' where the torus degenerates.
In fact, the degeneration locus can be identified with the configuration
of the $(N,N')$ web of 5-branes. In type IIA these correspond to webs of D6-branes which 
stretch along the 1-cycles of the torus according to the $(N,N')$ charges of the bound state, 
giving rise to a purely geometric M-theory background. 
 
One can further 
also consider more general webs of branes, which all give rise to 1/4 
BPS bound states. 
Using $SL(2,\mathbb{Z})$ transformation any junction of three branes in a 
web can be mapped to standard form, consisting of the intersection 
of a (1,0) brane, a
(0,1) brane plus one of type $(N,N')$ which together preserve 1/4 of supersymmetries
provided the angle of the $(N,N')$ brane is precisely the one
in the toric diagram of the degeneration locus (for more discussion see
\cite{Leung:1998tw}). Therefore we expect to find a large class of non-compact geometries 
with a metric of $spin(7)$ holonomy by defining degenerate fibrations over $\mathbb{P}^1$ (or 
its double cover) in terms of the respective brane web. Put another way, we compactify 
six-dimensional critical theories with 8 supercharges on the deformed conifold where the D6-branes 
wrap an $\mathbb{S}^3\times \mathbb{S}^1$ to get a three-dimensional theory with 2 supercharges. 
If this works generically, then any such theory should give rise to 
a (non-compact) $spin(7)$ manifold. 
Note the difference of this construction compared to the $spin(7)$
manifolds in \cite{Gomis:2001vk,Gukov:2001hf,Gauntlett:2001ur}, 
which are constructed by wrapping D6-branes over
supersymmetric four-cycles inside a $G_2$ manifold.

\section{M-theory on the $G_2$ and $spin(7)$ manifolds}
\setcounter{equation}{0}

In this chapter we add some further observations on the geometry of the M-theory 
backgrounds whose origin was described in the previous one.  

\subsection{The 1/2 BPS flux and M-theory on a $G_2$ manifold}

In this chapter we will embed the 1/2 supersymmetric 2-form
fluxes ({\it cases (i)} and {\it (ii)} in the previous section) 
through an $\mathbb{S}^2$ into a six-dimensional, non-compact
Calabi-Yau geometry. Since the background geometry preserves 1/4 of
supersymmetry, the total number of supercharges is 4. Therefore the
M-theory lift has to be described by a seven-dimensional space 
of $G_2$ holonomy.
We will consider as specific example the model
of \cite{Vafa:2000wi,Atiyah:2000zz},
 namely the non-compact
Calabi-Yau space $X_6$ is given by the resolved conifold
${\cal O}(-1)\oplus{\cal O}(-1)\longrightarrow {\mathbb P}^1$.
This is dual to the situation where the flux is replaced by D6-branes
being wrapped over the $\mathbb{S}^3$ of the deformed conifold $T^*\mathbb{S}^3$.

First consider the simplest situation  of having
$N$ units of RR 2-form 
flux through the $\mathbb{S}^2$ at finite radius ({\it case (i)} before). 
This flux induces a superpotential in four dimensions of the form
\begin{equation}
W=N~\Pi\, ,\qquad N=\int_{\mathbb{S}^2}H^{(2)}_R\, ,\qquad
\Pi=\int_{{\cal C}^{(4)}}(J\wedge J)\, \label{simplesup}
\end{equation}
where the integral which defines $\Pi$ goes over the non-compact
4-cycle ${\cal C}^{(4)}={\mathbb R}\times \widetilde{\mathbb{S}}^3$, being dual to the $\mathbb{S}^2$.
The M-theory lift is provided by replacing the six-dimensional
conifold geometry by a seven-dimensional
non-compact $G_2$ manifold $X_7$.
It is given by the spin bundle over $\mathbb{S}^3$ with topology
of ${\mathbb R}^4\times \mathbb{S}^3$. Asymptotically this $G_2$
manifold has the form of a cone whose base is topologically
$\widetilde{\mathbb{S}}^3\times \mathbb{S}^3$, represented as
\begin{equation}
(|z_1|^2+|z_2|^2)-(|\widetilde z_1|^2+|\widetilde z_2|^2)=V>0\, .\label{gtwo}
\end{equation}
Here the complex
coordinates $\widetilde z_1$, 
$\widetilde z_2$ parametrize $\widetilde{\mathbb{S}}^3\times 
\mathbb{R}_+$, whereas $z_1$, $z_2$ provide the coordinates of $\mathbb{S}^3$.

To get $N$ units of RR 2-form flux
the eleventh circle $\mathbb{S}^1_{11}$ is embedded
into $\mathbb{S}^3$. The discrete group ${\mathbb Z}_N\subset U(1)$ has
a non-singular action on $\mathbb{S}^3$,
since the volume of this sphere is nowhere vanishing.
The corresponding
 $\mathbb{S}^1_{11}$ fibration is the quotient of the Hopf fibration by
${\mathbb Z}_N$, namely the  space $\Sigma_3$ of the previous
chapter.
It defines $X_7$ as being 
the spin bundle over $\Sigma_3$ which is at the same time
a $\mathbb{S}^1_{11}$ fibration but now
over a six-dimensional base $B_6$:
\begin{eqnarray}
&{~}& {\mathbb R}^4\longrightarrow 
X_7\longrightarrow\Sigma_3\, , \quad 
\mathbb{S}^1_{11}\longrightarrow X_7\longrightarrow B_6\, .
\end{eqnarray}
The metric of this non-compact $G_2$ manifold is known 
\cite{Atiyah:2000zz,Brandhuber:2001yi,aw} 
and takes the
following form (for the case $N=1$):
\begin{equation}
X_7:\qquad ds^2=
dr^2+a(r)^2\sum_{a=1}^3(\widetilde\sigma_a)^2+
b(r)^2\sum_{a=1}^2(\sigma_a-{1\over 2}
\widetilde\sigma_a)^2+c(r)^2(\sigma_3-{1\over 2}\widetilde\sigma_3)^2
\, .\label{metricgtwo}
\end{equation}
Here $\sigma_a$ and $\widetilde\sigma_a$ are the left-invariant 1-forms
of the two three-spheres $\mathbb{S}^3$ and $\widetilde{\mathbb{S}}^3$;
$a(r)$, $b(r)$ and $c(r)$ are $r$-depedent functions whose forms
are dictated by the requirement of having $G_2$ holonomy.
Asymptotically, for $r\rightarrow\infty$, these functions
behave like $a(r),b(r)\rightarrow r$ and $c(r)\rightarrow{\rm constant}$.
This asymptotic behaviour of $c(r)$ ensures that one has a constant
$U(1)$ fibre at infinity, which corresponds to the differential
$\sigma_3$ \cite{Brandhuber:2001yi,aw}.
One can reduce the theory along the circle $\mathbb{S}^1_{11}$, corresponding
to setting $c(r)=0$.
Then one obtains the metric of the small
resolution of the conifold, ${\cal O}(-1)\oplus{\cal O}(-1)\longrightarrow
{\mathbb P}^1$, with  a non-trivial
RR 1-form potential $A_\mu^{11}$ in addition, which corresponds
to $N$ units of RR two form flux on $\mathbb{S}^2$.

As explained already in the dual type IIA brane picture one considers
$N$ D6-branes wrapped around the $\mathbb{S}^3$ of the deformed conifold geometry,
which are lifted in M-theory to $N$ ${\cal KK}$ monopoles given in terms
of an $A_{N-1}$ singularity transversal to the $\mathbb{S}^3$. In the $G_2$
example this is achieved by constructing an ${\mathbb R}^3$
bundle over the four-dimensional
$A_{N-1}$ singularity, which means that one is
considering instead the singular
quotient of $\widetilde{\mathbb{S}}^3$ by ${\mathbb Z}_N$.
Then it is clear that the two cases, the
wrapped brane picture and the flux picture, are  related by the exchange
of the singular quotient by the non-singular one. Geometrically this
can be realized by performing the exchange
$V\rightarrow -V$ in eq.(\ref{gtwo}), which is just the flop transition in the
$G_2$ geometry.

Let us now briefly switch to {\it case (ii)} where we have $N$ units of 
RR 2-form flux plux $N'$ units of NSNS 2-form flux through the same
$\mathbb{S}^2$ of the resolved conifold, where the fluxes preserve still 1/2
supersymmetry. So we now add the circle $\mathbb{S}^1_3$ in the third direction,
such that we consider M-theory on an eight-dimensional space $X_8$.
This space is now a spin bundle over $\Sigma_4$. 
Due to the four conserved supercharges, the topology of $X_8$ must be
given by $X_7\times \mathbb{S}^1_3$, where, as before, $X_7$ has $G_2$
holonomy. The metric of this space can be obtained from the $G_2$ metric
eq.(\ref{metricgtwo}) by a simple $SL(2,\mathbb{Z})$ transformation, as
described for the space $\Sigma_4$ (see eq.(\ref{metricsigmafour})).
Namely we add just the new coordinate $\psi'$, and then we obtain a 
$SL(2,{\mathbb Z})$ family
of eight-dimensional metrics, labelled by the two integers $N,N'$, of
the following form\footnote{It is conceivable that the choice of the two
integers $N,N'$, which form an $SL(2,{\mathbb Z})$ doublet, is related to
the framing ambiguity discussed in \cite{Aganagic:2001nx,Aganagic:2001jm};
we like to thank Sergei Gukov for discussions on this point.}
\begin{equation}
X_8:\qquad ds^2=
dr^2+a(r)^2\sum_{a=1}^3(\widetilde\sigma_a)^2+
b(r)^2\sum_{a=1}^2(\sigma_a'-{1\over 2}
\widetilde\sigma_a)^2+c(r)^2(\sigma_3'-{1\over 2}\widetilde\sigma_3)^2
+c'(r)^2(\sigma_4')^2\, .\label{metricgtwoprime}
\end{equation}
where the $\sigma_a'$ are given by eq.(\ref{sigmaprimes}).
The functions $a(r),b(r),c(r),c'(r)$ for large $r$ again have the asymtotics
$a(r),b(r)\rightarrow r$, $c(r),c'(r)\rightarrow {\rm constant}$,
which exhibits the last two terms in (\ref{metricgtwoprime}) as
the finite ${\mathbb T}^2$ fibre at infinity.

M-theory on this background space $X_8$ leads to  a three-dimensional
field theory with four supercharges (${\cal N}=2$ supersymmetry).
Following the analogous derivation of the ${\cal N}=1$ superpotential
in four dimension, the corresponding superpotential in three dimensions
is given by the following expression:
\begin{equation}
W=(R_{11}N+iR_3N')~\Pi=R_{11}(N+\tau N')~\Pi\, . 
\label{suporns}
\end{equation}
This superpotential is fully covariant under the S-duality
transformations 
\be 
\tau\rightarrow{a\tau+b\over c\tau+d},\quad 
\pmatrix{N\cr N'}\rightarrow
\pmatrix{a&b\cr c&d}\pmatrix{N\cr N'}\, .
\ee
Via $SL(2,\mathbb{Z})$ transformation it can be brought back into
the form eq.(\ref{simplesup}), which means that in this case the
three-dimensional superpotential can be obtained from the
four-dimensional one by simple dimensional reduction on a circle.

\subsection{The 1/4 BPS flux and M-theory on a $spin(7)$ manifold}

Now we discuss the M-theory compactification on an eight-dimensional
(non-compact) space $X_8$, which corresponds to the 1/4 BPS situation.
The holonomy is easily classified in terms of the number of 
conserved supercharges, which we have employed already at various instances. 
For $N=N'=0$, $X_8$ is simply the direct product $X_6\times \mathbb{T}^2$, where
$X_6$ is a six-dimensional Calabi-Yau space with $SU(3)$
holonomy, corresponding to eight unbroken supercharges.
In the case of $N\neq0$, $N'=0$, or vice versa, 
the flux breaks half of the supersymmetry, so
together with the Calabi-Yau background 1/8 of SUSY is preserved.
Then $X_8$ is the direct product of an $G_2$ manifold $X_7$ times
an $\mathbb{S}^1$: $X_8=X_7\times \mathbb{S}^1$. 
Finally, for the 1/4 supersymmetric fluxes only 1/16
supersymmetries are preserved. Therefore the holonomy $H$ of $X_8$
has to be given by $spin(7)$. Now
the torus $\mathbb{T}_{3,11}^2=\mathbb{S}^1_{11}\times \mathbb{S}^1_3$
is non-trivially fibred over a six-dimensional base and circles degenerate on 
codimension one loci in the manner described above. 
$H$ must actually contain $SU(3)$ as well as $G_2$ as subgroups, according to 
the subsectors with enhanced supersymmetry: 
\begin{equation}
SU(3)\longrightarrow G_2\longrightarrow spin(7)\, .
\end{equation}

Let us now try to characterize the non-compact $spin(7)$ manifold
which arises putting the RR plus the NSNS 2-form flux through
the $\mathbb{S}^2$ of the resolved conifold geometry. In analogy to the
$G_2$ manifold considered before $X_8$ is a fibre 
bundle now over the four-dimensional space 
$\Sigma_4$, which is itself a torus fibration with degenerate fibres.
As discussed, a particular example of this kind is given by $\Sigma_4=
\widetilde{{\mathbb CP}^2}$.
Roughly spoken, $X_8$ can be constructed by fibering another circle
over the seven-dimensional  $G_2$ manifold.
More precisely,
$X_8$ can be viewed as a $\mathbb{S}^1_3$ fibrations over a seven-dimensional base
space $B_7$
which is the  non-Ricci flat relative of the 
$G_2$ manifold $X_7$ considered before:
\begin{equation}
{\mathbb R}^4\longrightarrow
X_8\longrightarrow \Sigma_4(\widetilde{{\mathbb CP}^2}) \quad{\rm or}\quad
\mathbb{S}_3^1\longrightarrow X_8\longrightarrow B_7\, .
\end{equation}
We can also 
represent $X_8$ as a torus fibration 
over $B_6$: 
\begin{equation}
\mathbb{T}^2_{3,11}\longrightarrow X_8\longrightarrow B_6\, , 
\end{equation}
where the torus fibration is however not smooth.

The relation between the fibration structures of the $G_2$ manifold
$X_7$ and the $spin(7)$ manifolds $X_8$ can be nicely
summarized by the following diagrams:

\parbox{20cm}{
\hspace{1.5cm}
\parbox{6cm}{
\begin{eqnarray} \nonumber 
&{}&\hskip0.5cm{\mathbb R}^4\hskip1.2cm {\mathbb R}^4\nonumber\\
&{}&\hskip0.5cm\downarrow\hskip1.5cm\downarrow\nonumber\\
\mathbb{S}_{11}^1\,\,\,\, \longrightarrow &{}&X_7\,\,\,\, \longrightarrow\,\,\,\, 
B_6\nonumber\\
&{}&\hskip0.5cm\downarrow\hskip1.5cm\downarrow\nonumber\\
\mathbb{S}_{11}^1\,\,\,\, \longrightarrow &{}&\Sigma_3\,\,\,\, \longrightarrow\,\,\,\,
\mathbb{P}^1\nonumber
\end{eqnarray}
}
\hspace{1cm}
\parbox{6cm}{
\begin{eqnarray}
&{}&\hskip0.5cm{\mathbb R}^4\hskip1.2cm {\mathbb R}^4\nonumber\\
&{}&\hskip0.5cm\downarrow\hskip1.5cm\downarrow\nonumber\\
\mathbb{S}_3^1\,\,\,\, \longrightarrow &{}&X_8\,\,\,\, \longrightarrow\,\,\,\, 
B_7\nonumber\\
&{}&\hskip0.5cm\downarrow\hskip1.5cm\downarrow\nonumber\\
\mathbb{S}_3^1\,\,\,\, \longrightarrow &{}&\Sigma_4\,\,\,\, \longrightarrow\,\,\,\,
\Sigma_3\nonumber
\end{eqnarray}
}
}

Now let us come to the construction of the metric of the $spin(7)$
manifold $X_8$. 
Recall that we propose a flux/brane setting, 
where the $\Sigma_4$ 
is not yet present in the $G_2$ manifold but is only build up 
by fibering the $\mathbb{S}^1_{11}$ respectively  $\mathbb{S}^1_3$ 
over the $\mathbb{P}^1$ in the parent 
$G_2$ space $\mathbb{R}^4 \times \mathbb{S}^3$. 
In the simplest case, we would expect an 
asymptotic metric of the form 
\be \label{asympt} 
ds^2 \longrightarrow \underbrace{dr^2 + r^2 
\big( \sigma_1^2 + \sigma_2^2 + \sigma_3^2 \big)}
_{\mathbb{R}^4} + \underbrace{r^2 
\big( \widetilde\Sigma_1^2 + \widetilde\Sigma_2^2 \big)}
_{{\rm Base}\ \mathbb{P}^1} +
\underbrace{\big( \widetilde\nu_1^2 + \widetilde\nu_2^2 \big)}
_{\rm Fibre\ \mathbb{T}^2} , 
\ee
with a finite radius for the fibre torus at infinity. 
Depending on how one takes the limit $r\rightarrow \infty$, i.e. 
at which 
point in the base, one obtains a different copy of the fibre torus at 
infinity, with a different circle staying finite. This respective circle 
than takes over the role of the type IIA coupling constant in the  
appropriate ten-dimensional vacuum.   
On the other hand, at small radius $r\rightarrow 0$, 
at least after some proper 
shifting of $r$, the metric should collapse to the singular orbit 
$\Sigma_4$:
\be
ds^2 \longrightarrow dr^2 +{\rm const}
\left( \widetilde\Sigma_1^2 + \widetilde\Sigma_2^2 + 
\widetilde\nu_1^2 + \widetilde\nu_2^2 \right) \, .
\ee

Let us now consider more closely the case, where 
$\Sigma_4=\widetilde{{\mathbb CP}^2}$.
As it is well known, one can construct metrics on an ${\mathbb R}^4$ bundle
over ${\mathbb CP}^2$ which indeed possess $spin(7)$ holonomy.
The most general form of such metrics considered so far is  
\cite{Cvetic:2001zx,Kanno:2001xh,Gukov:2001hf}:
\begin{equation} 
ds^2=dr^2+f^2(r)\lambda^2+a^2(r)(\mu_1^2+\mu_2^2)+
b^2(r)(\Sigma_1^2+\Sigma_2^2)+c^2(r)(\nu_1^2+\nu_2^2)\, .
\label{spinsevenmetric}
\end{equation}
The $\nu_i$ and $\Sigma_i$ are the differentials on $\mathbb{C}P^2$ while $dr$, $\lambda$ 
and $\mu_i$ parametrize the $\mathbb{R}^4$. The formalism is 
based on a coset construction 
of the manifold, and thus the forms $\nu_i$, $\Sigma_i$, $\mu_i$ and 
$\lambda$ together with one more Cartan generator satisfy 
the algebra of $SU(3)$, the latter one 
being removed in the coset $SU(3)/U(1)$.     
The important feature of this cohomogeneity one ansatz 
is the fact that the coefficient functions only 
depend on a radial coordinate $r$ and 
not on the angular coordinates. 
Only a very few explicit solutions are in fact known. 
The asymptotic behaviour for $r\rightarrow 
\infty$ of the ``squashed solution'' is similar to that 
known for Taub-NUT spaces, 
where one circle stays at finite radius, 
\be 
ds^2 \longrightarrow dr^2 + \frac{9\rho^2}{4} \lambda^2 + r^2 
\left( \mu_1^2 + \mu_2^2 + \Sigma_1^2 + \Sigma_2^2 
+ \nu_1^2 + \nu_2^2 \right) 
\ee
with some constant $\rho$. Here, the $U(1)$ direction that corresponds to the 
eleventh direction is identified with the $SU(3)/U(1)$ Cartan 
generator $\lambda$. 
This is interpreted to describe 
the presence of a IIA D6-brane wrapped on the supersymmetric 4-cycle 
of the $G_2$ 
manifold $\Lambda^- \mathbb{C}P^2$ which is topologically 
$\mathbb{C}P^2 \times\mathbb{R}^3$. 
In other words, in M-theory
the ${\cal KK}$-monopole  wraps on the $\mathbb{C}P^2$ and the 
eleventh circle, 
given by $\lambda$,  
being fibered non-trivially, but still 
approaches constant radius at infinity and thus supports a type IIA 
string vacuum with a finite coupling constant. 

This configuration 
is indeed not quite what we have in mind. 
In order to realize the asymptotic behaviour (\ref{asympt}) in terms of
the splitting of base and 
fibre in the metric of $\mathbb{C}P^2$ (see eq.({\ref{fubini})),
one is now tempted to identify  
\beqa  
ds_{\rm FS}^2=\Sigma_1^2+\Sigma_2^2+\nu_1^2+\nu_2^2 
=\widetilde\Sigma_1^2+\widetilde\Sigma_2^2+
\widetilde\nu_1^2+\widetilde\nu_2^2 . 
\eeqa
The choice of the Vielbein components $\widetilde{\Sigma}$ and 
$\widetilde{\nu}$ 
is of course not unique. 
However analyzing the differential system in \cite{Gukov:2001hf}
which ensures $spin(7)$ holonmy, 
one can show that the function $c(r)$ in eq.(\ref{spinsevenmetric})
can never approach a constant for $r\rightarrow\infty$.
Also one needs to take care that the differentials  $\widetilde\Sigma_i$
and $\widetilde\nu_i$
can never satisfy the algebra of $SU(3)$ together 
with the generators $\mu_1$ and $\mu_2$, as do $\Sigma_i$ and $\nu_i$. 
When the metric on the base does 
not depend on the coordinates of the fibre at large radius $r$, 
this signals a breaking 
of the $SU(3)$ invariance of the metric asymptotically. 
Thus, the $spin(7)$ manifold we propose 
to exist on physical grounds does seem to fit easily into the 
known classes of examples, 
constructed from the coset formalism. 

As we will discuss in the appendix,
one finds that the asymptotic restrictions cannot be solved within 
the cohomogeneity one ansatz, 
if one tries to stick to the differentials (\ref{tildediff}).  
Hence, the expectation (\ref{asympt}) may even be too simple and 
one should allow for completely 
independent coefficients for the differentials on $\mathbb{S}_{11}^1$ and $\mathbb{S}^1_3$ 
in the metric, which then may 
depend on $r$ as well as on the coordinates of the base space. 
For further explanations the reader is referred to the appendix. 

In any case, the Fubini Study metric on $\widetilde{\mathbb{C}P^2}$ 
allows a split into base 
$\mathbb{P}^1$ and a fibre $\mathbb{T}^2$, and a solution with asymptotics  
of the type described by (\ref{asympt}) would presumably 
reproduce the M-theory dynamics 
of the type II vacua with respective fluxes or branes 
considered earlier. 
The explicit metric still remains to be found, its existence, in the absence 
of any mathematical existence proof, being inferred only on physical grounds.

\section{Some remarks on higher $n$-form fluxes}
\setcounter{equation}{0}

In this section Vafa likes to add some remarks on the possible
explanation of higher NS-fluxes, such as NSNS 4-flux and NSNS 6-flux.

\subsection{4-form flux}

As already discussed in the introduction, in the mirror type IIB
picture the four-dimensional superpotential due to 3-fluxes has the form
\begin{equation}
W=
N~\Pi+\alpha~t\, ,\label{vafasupa}
\end{equation}
where the gauge coupling constant 
$\alpha=\int_{{\cal C}^{(3)}}\tau H^{(3)}_{NS}$ is the NSNS 3-form
flux over the non-compact cycle ${\cal C}^{(3)}$, and 
the gaugino condensate $t={\rm Tr}\, W^2=\int_{\mathbb{S}^3}\Omega$
is the modulus of the dual 3-sphere $\mathbb{S}^3$.
Now in type IIA the origin of the gaugino condensate term
in the superpotential is somehow mysterious, and in fact it was argued in
\cite{Vafa:2000wi} that it is related to the NSNS 4-form flux.
Namely turning on the 
RR 2-form flux implies that $X_6$ is actually not anymore a Calabi-Yau
space but replaced by the six-dimensional base
$B_6$ of the
$G_2$ manifold.
However the holomorphic three-form $\Omega$ of
$X_6$ is no longer closed on $B_6$, $\Omega$ is not anymore
annihilated by the operator $\bar\partial$. Instead of $\bar\partial$
Vafa \cite{Vafa:2000wi} introduces a new operator
$\bar{\cal D}=\bar\partial+A\partial$, where $A$ is an anti-holomorphic
1-form taking values in the tangent bundle, $A\in\Lambda^{(0,1)}(X,TX)$.
Squaring this operator leads to an antiholomorphic 2-form taking values
in the tangent bundle as well, given as
\begin{equation}
{\cal F}=\bar \partial A+\lbrack A,A\rbrack\, ,
\quad {\cal F}=f_{\bar i\bar j}~d\bar z_i\wedge d\bar z_j
\otimes dz^k\in\Lambda^{(0,2)}(X,TX)\, .
\end{equation}
Finally, multiplying ${\cal F}$ with the holomorphic (3,0)-form $\Omega$
one obtains a (2,2)-form, which we like to identify with ${\rm Im}
~\tau H^{(4)}_{NS}$:
\begin{equation}
{\rm Im}~
\tau H^{(4)}_{NS}=f_{\bar i\bar j}~d\bar z_i\wedge d\bar z_j\otimes dz^k~
\Omega_{ijk}~dz_i\wedge dz_j\wedge dz_k\in\Lambda^{(2,2)}(X)\, .
\end{equation}
This (2,2)-form can be used to define
the imaginary part of $\alpha$.
Going to M-theory it can be then further argued 
\cite{Vafa:2000wi,Cachazo:2001jy}
that the NSNS 4-form flux $\alpha$ corresponds
after appropriate regularization  to the holomorphic
volume of the 3-cycle $\tilde \mathbb{S}^3$, i.e.
\begin{equation}
\alpha=
\int_{\widetilde{\mathbb{S}}^3}(C^{(3)}+i\Omega)
=\theta+i{\rho\over g_s}\, ,\label{rhotheta}
\end{equation}
where $C^{(3)}$ is the M-theory 3-form potential.

The conclusion of this argument is that 
the RR 2-form flux is necessarily turned on together with 
NSNS 4-form flux, which actually arises as a violation of $\bar{\partial}\Omega=0$. 
This can be also seen in the effective superpotential
eq.(\ref{vafasupa}). Here the conditions for finding supersymmetric
groundstates, $W=dW=0$, turn into the following relation \cite{Vafa:2000wi}:
\begin{equation}
(e^t-1)^N=a\exp(-\alpha)\, .
\end{equation}
So one sees that unbroken supersymmetry intimately ties together
the RR 2-form flux with the NSNS 4-form flux.

In the following let us try to give to  
the real part $\theta$ in eq.(\ref{rhotheta}) 
a 4-flux interpretation. 
Since the RR 4-form field strength $H^{(4)}_R$ is present
in type IIA, we can
immediately integrate it over the non-compact
four-cycle ${\cal C}^{(4)}={\mathbb R}_+\times
\widetilde{\mathbb{S}}^3$ of the deformed conifold,
which is dual to the $\mathbb{S}^2$ considered before.
In M-theory, this originates just from the standard $G$-flux, i.e.
it is given by
the intergral of the eleven-dimensional field strength $G^{(4)}=dC^{(3)}$,
integrated over the same ${\cal C}^{(4)}$ inside $X_7$.
This integration does not involve the eleventh circle $\mathbb{S}^1_{11}$,
and we can obtain the following Ramond contribution to the(four-dimensional)
superpotential eq.(\ref{supofourflux}):
\begin{equation} 
{\rm Re}~W=\int_{{\mathbb R}_+\times
\widetilde{\mathbb{S}}^3}G^{(4)}~\int_{\mathbb{S}^2}J=t~\int_{\widetilde{\mathbb{S}}^3}
C^{(3)}=\theta ~t\, , \quad t=\int_{\mathbb{S}^2}J \, .
\label{suporfour}
\end{equation}
This indeed agrees with eq.(\ref{rhotheta}).


\subsection{6-form flux}

In this section we want to briefly discuss a contribution to the
three-dimensional
superpotential of the form
\begin{equation}
W=\int_{X_6}H^{(6)}_R+i\int_{X_6}H^{(6)}_{NS}=N_0^R+{i\over g_s}
N_0^{NS}\, .\label{suposix}
\end{equation}
The RR part of this superpotential simply follows from the integration
of the 6-form
field strength $H^{(6)}_R$ which is present in the type IIA spectrum. 
One can lift this to M-theory by considering the 7-form field strength 
$H^{(7)}$, dual to the field strength $G_4=dC_3$, where
$C_3$ is the 3-form potential of M-theory. 
Specifically, we assume in this section that M-theory is
compactified on $X_6\times \mathbb{S}^1_{11}\times \mathbb{S}^1_3$, where $X_6$ is
Calabi-Yau, which means that we assume for
simplicity that there are no other fluxes turned on. 
Then the RR 6-form flux is given in terms of the integral of
$H^{(7)}$, where six of the indices of $H^{(7)}$ are on
$X_6$ and the last index is on $\mathbb{S}^1_{11}$:
\begin{equation}
\int_{X_6}H^{(6)}_R=\int_{X_6\times \mathbb{S}^1_{11}}H^{(7)}=R_{11}N_0^R\, .
\label{suposixr}
\end{equation}

On the other hand, the NSNS 6-form flux is obtained by replacing
$\mathbb{S}^1_{11}$ by $\mathbb{S}^1_3$, namely by integrating $H^{(7)}$ 
with six
indices on $X_6$ and one index on $\mathbb{S}^1_3$:
\begin{equation}
i\int_{X_6}H^{(6)}_{NS}=i\int_{X_6\times \mathbb{S}^1_{3}}H^{(7)}=
iR_{3}N_0^{NS}\, .
\label{suposixns}
\end{equation}
We see that the sum of the superpotentials eqs.(\ref{suposixr}) and
(\ref{suposixns}) agrees with eq.(\ref{suposix}) upon rescaling
with $1/R_{11}$.

\section{Summary}
\setcounter{equation}{0}

Let us summarize the different cases of flux backgrounds we have
considered in this paper.
$N$ units of RR 2-form flux (together with a NSNS 4-form flux) 
on the conifold preserve four
supercharges and leads to four-dimensional
${\cal N}=1$ $U(N)$ gauge theory with a flux induced superpotential.
By reducing further to three dimensions, the type IIB $SL(2,\mathbb{Z})$ 
duality can be made manifest by a doublet of fluxes $(N\ N')$,
which corresponds to $N$ units of Ramond 2-form flux together
with $N'$ units of NSNS 2-form flux. For 1/2 supersymmetric fluxes this 
situation is described in M-theory by a $SL(2,\mathbb{Z})$
family of (eight-dimensional) metrics of $G_2$ holonomy of spaces of
topology $X_7\times S^1$.
This setting has a dual brane description in terms of 
1/2 BPS bound states of $N$ D6-branes plus $N'$ Kalazu-Klein
monopoles wrapped 
on sLag cycles within the Calabi-Yau 3-fold. Space-time
supersymmetry will be further reduced by half, if one 
simultaneously turns on RR as well as NSNS 2-form flux
on the conifold times a torus in a particular and singular fashion, 
that is related 
to a dual 1/4 BPS brane web state. This configuration leads to a field theory
with minimal ${\cal N}=1$ supersymmetry in three dimensions so that 
these fluxes realize partial supersymmetry
breaking from ${\cal N}=4$ to ${\cal N}=1$ supersymmetry. 
Geometrically, they are described by M-theory on a $spin(7)$
manifold.\footnote{Other geometric realizations of three-dimensional
field theories with ${\cal N}=1$ supersymmetry are provided by
the heterotic string on a $G_2$ manifold or by S-theory on U-manifolds
(certain $\mathbb{T}^3\times \mathbb{T}^2$ fibred Calabi-Yau five-folds 
\cite{Kumar:1997zx,Liu:1998mb,Curio:1998bv}).}
It would be interesting to explore further the geometric
understanding of non-perturbative effects in
the corresponding ${\cal N}=1$ supersymmetric field theories 
\cite{Affleck:1982as,Katz:1997th}.
Following \cite{Witten:1995rz}
one could try to 
relate these ${\cal N}=1$ supersymmetric theories in three
dimensions to non-supersymmetric theories in four-dimensions
with vanishing cosmological constant.

\begin{appendix}

\section{$Spin(7)$ metric ansatz}  
\setcounter{equation}{0}

In this appendix we employ the formalism which has been used 
for the construction of all the known 
explicit metrics with $spin(7)$ holonomy.
First recall that there are three different non-compact $G_2$ manifolds
$X_7$, each given as a cone on some compact six-dimensional space $Y_6$ \cite{aw}:
\begin{equation}
({\rm A}):\,\,\, Y_6={\mathbb CP}^3,\quad ({\rm B}):\,\,\, 
Y_6=SU(3)/(U(1)\times U(1)),\quad
({\rm C}):\, \,\, Y_6=\mathbb{S}^3\times \mathbb{S}^3\, .
\end{equation} 
Asymptotically for large $r$
the metric takes the general form
\begin{equation}
ds^2=dr^2+r^2d\Omega_6^2\, ,
\end{equation}
where $d\Omega_6^2$ is the metric on $Y_6$.
Topologically spaces (A) and (B) are some ${\mathbb R}^3$ bundle
over $\mathbb{S}^4$ respectively  ${\mathbb CP}^2$, whereas space (C) is a ${\mathbb R}^4$
bundle over $\mathbb{S}^3$.
All three spaces possess a $U(1)$ fibration
\begin{equation}
\mathbb{S}^1\longrightarrow  X_7\longrightarrow B_6\, ,
\end{equation}
which means that one can squash the metrics in such a way that
the remainder of the sqashed
metric asymptotically
is a cone over a reduced five-dimensional space $Z_5$:
\begin{equation}
ds^2=dr^2+\rho\lambda^2+r^2d\Omega_5^2\, .
\end{equation}
Here $\rho$
is a constant, $\lambda$ is the Vielbein corresponding 
to the $U(1)$ fibre, with finite radius at infinity, and
$d\Omega_5^2$ is the metric of $Z_5$.
For example, for case (C), $Z_5=T^{1,1}=(SU(2)\times SU(2))/U(1)_D$, the
base of the cone of the conifold.

Roughly speaking, we would now like fibre  another $U(1)$ over each of the
three $G_2$ manifolds $X_7$  (A) -- (C).
The resulting spaces are given by non-compact $spin(7)$ manifolds
$X_8$, which are themselves asymptotically cones over a seven-dimensional
space $Y_7$. Cases (A) and (B) were discussed e.g. in 
\cite{Cvetic:2001zx,Kanno:2001xh,Gukov:2001hf}:
\begin{equation}
({\rm A}):\, \,\, Y_7=SO(5)/SO(3),\quad ({\rm B}):\, \,\, Y_7=SU(3)/U(1)
\, .
\end{equation}
The 
spaces (A) and (B) are ${\mathbb R}^4$ bundles over $\mathbb{S}^4$ or 
${\mathbb CP}^2$, respectively.
Again one can squash the corresponding metrics in order to exhibit
the $U(1)$ fibration structure, and then the remainder of the
squashed metrics is a cone over a six-dimensional space $Z_6$.
For case (A), this space is given $Z_6={\mathbb CP}^3$.
In terms of branes, this $U(1)$ fibration can be understood
by wrapping D6-branes around the co-associative 
$\mathbb{S}^4$ cycle in the
parent $G_2$ manifold (A).
For case (B) with
metric eq.(\ref{spinsevenmetric}),
being described in the main part of the text,
$Z_6=SU(3)/(U(1)\times U(1))$, which corresponds to wrapping 
D6-branes around the ${\mathbb CP}^2$ inside the corresponding
$G_2$ manifold (B).

The manifold we are after has been characterized above as arising 
from the degenerate fibration of a torus over an $\mathbb{S}^2$ 
and at the same time allowing two limiting cases where it appears as 
adding a circle fibre to a $G_2$ 
manifold, which itself is defined by a quotient of a Hopf fibration. 
Thus, we are mainly interested in the construction of a $spin(7)$
manifold (C) which is the close relative of the 
corresponding $G_2$ manifold (C). Replacing  $\mathbb{S}^3$ by $\Sigma_4$,
the $spin(7)$ manifold should be a ${\mathbb R}^4$ bundle over
$\Sigma_4$. Asymptotically this corresponds to a cone
on $\mathbb{S}^3\times \Sigma_4$. Upon `dividing' by the ${\mathbb T}^2$ fibre
$\mathbb{S}^1_{11}\times \mathbb{S}^1_3$ the remaining metric is again a cone on
$T^{1,1}$.
Let us now assume that $\Sigma_4={\mathbb CP}^2=SU(3)/(SU(2)\times U(1))$ 
(respectively $\Sigma_4=\widetilde{{\mathbb CP}^2}$). 
In order to construct the metric of this
space (C), it would be quite natural to start with an ansatz which
uses the $SU(2)$ differential $\sigma_a$ and the differentials
$\widetilde\Sigma_i$, $\widetilde\nu_i$ of the coset 
$SU(3)/(SU(2)\times U(1))$:
\beqa \label{bestansatz}
ds^2 &=& dr^2+a^2(r)(\sigma_1^2+\sigma_2^2+\sigma_3^2)\non &&  +
b^2(r)\left( (\widetilde\Sigma_1-{\gamma\over 2}\sigma_1)^2
+(\widetilde\Sigma_2-{\gamma\over 2}\sigma_2)^2\right)
+c^2(r)\left( (\widetilde\nu_1-{\beta\over 2}\sigma_3)^2+
(\widetilde\nu_2-{\beta\over 2}\sigma_3)^2\right)\, .
\eeqa
We take $\beta$ and $\gamma$ as two free real parameters to play with.
This also would potentially reflect the asymptotic behaviour in eq.(\ref{asympt}).
However the difficulty with this ansatz is that the Vielbeins
of ${\mathbb CP}^2$ do not build a closed algebra. 
This can be immediately seen from the Maurer-Cartan
equation of $SU(3)$,
\begin{equation}
de^a=-{1\over 2}f^a_{bi}~e^b\wedge e^i-{1\over 2}f^a_{bc}~e^b\wedge e^c\, ,
\end{equation}
where the algebra of the coset generators $e^a$ also  involves
the generators $e^i$ of the $SU(2)\times U(1)$ isotropy group.
It follows that the construction of the corresponding
metric would go beyond the cohomogeneity one ansatz, and is therefore
very difficult to handle explicitly.

As  an alternative, one may consider an $SU(3)$ invariant ansatz,
where the manifold arises from the coset $SU(3)/U(1)$. 
This would not quite be the structure we 
expect to emerge, because it does not display the pattern of 
the two $G_2$ manifolds reached by taking certain limits
in an obvious way, i.e. 
going to certain points in the base of $\mathbb{C}P^2$. But, on the 
other hand, this manifold explicitly involves a $\mathbb{C}P^2$.  
For instance, one may want to use an ansatz (in the notation of 
\cite{Gukov:2001hf}, where $\lambda, \mu_1, \mu_2, \nu_1,\nu_2, \Sigma_1$ 
and $\Sigma_2$ are coordinates on the coset) 
\beqa \label{neueransatz}
ds^2 &=& dr^2+f^2(r)\lambda^2+a^2(r)(\sigma_1^2+\sigma_2^2) \non && +
b^2(r)\left( (\Sigma_1-{\gamma\over 2}\sigma_1)^2+(\Sigma_2-
{\gamma\over 2}\sigma_2)^2\right)
+c^2(r)\left( (\nu_1-{\beta\over 2}\lambda)^2+(\nu_2-
{\beta\over 2}\lambda)^2 \right)\, .
\eeqa 
But it turns out that this does not respect the $U(1)$ symmetry, 
thus is not defined properly on $SU(3)/U(1)$.\footnote{We would like 
to thank V.~Braun who has drawn our attention to this point which was missed 
in an earlier version of this paper.}

So in conclusion, 
we should better stick to the coordinates which realize 
the splitting in the metric into base and fibre within $\mathbb{C}P^2$, 
as in (\ref{splitting}). But then, as mentioned already, 
the differentials $\widetilde \Sigma$ and 
$\widetilde\nu$ do not satisfy algebraic relations anymore. Therefore, such an 
ansatz, which is what we would really like to carry through, 
is beyond the scope of present techniques, unfortunately.      

\end{appendix}

\section*{Acknowledgements}

We would like to thank Mina Aganagic, Volker Braun, Sergei Gukov,
Albrecht Klemm, 
Tony Pantev and Ashoke Sen
for valuable discussions. D.L. is especially grateful to Jan Louis who was
involved in an early stage of this project. This work is supported by the 
Deutsche Forschungsgemeinschaft (DFG), by the European 
Commission RTN programe HPRN-CT-2000-00131 and by GIF - the German-Israeli 
Foundation for Scientific Research.  
In addition this research was supported in part by the National Science
Foundation under Grant No. PHY99-07949 through the Institute for
Theoretical Physics in Santa Barbara. D.L. thanks the ITP
for the hospitality during part of this work.

\end{document}